\documentstyle[11pt]{article}
\begin{document}
\begin{center}

\bf The H$_{\alpha}$, H$_{\beta}$\rm\, and H$_{\delta}$,\rm\, in
the spectrum of RR Tauri\\
 \vspace{8mm}
 \large
 V.I. Kardopolov,  L.N. Kondratyeva

 \vspace{4mm}

{\it Fessenkov Astrophysical Institute, 480020, Almaty, Kazakhstan}\\

    \vspace{4mm}

\end{center}
\vspace{4mm} \small \noindent The results of spectroscopic study
of Ae Herbig star RR Tau are reported. The observations were
carried out with a moderate resolution. In accordance with the
presented data the H$_{\alpha}$\rm\,line exhibits a double-peaked
emission profile. The blue component is weaker than or equal to
the red one. Variations of the H$_{\alpha}$\rm\, equivalent width
by a factor of 4 - 5 were found. The structure of this line also
varies. On JD 2447148 a red absorption reversal of the
H$_{\alpha}$ profile was observed. The H$_{\beta}$\rm\,  spectral
line exhibited both emission and absorption. On JD 2449664 the
single-peak emission shifted to the short wavelength range of the
underlying absorption  line and the red wing of the stellar
 line looked like a red-shifted absorption component of medium strength.
 The H$_{\delta}$\rm\, line was in absorption.\\
\vspace{6ex} \noindent{KEY WORDS\,}\,\rm\mbox{Young stars, Herbig
Ae/Be stars, spectroscopic observations}

     \normalsize
\section{INTRODUCTION }
The term Herbig Ae/Be  (HAEBE) stars owes its origin to the idea
that it is possible to observe hot pre-main-sequence stars
\cite{her60}. Calculations of early stellar evolution allow one to
predict the properties of HAEBE stars in detail, to  interpret (to
a first approximation) environment effects, and to outline an
optimum programme of investigations \cite{str72}. The well-known
peculiarity of the continuum activity is the blueing effect which
is observed in the deep minima of those HAEBE stars which exhibit
variations of a few magnitudes in the amplitude of the light, (see
for example \cite{kar85,kil85,kol77,sha85,za73}). Recently
simultaneous polarimetric and photometric measurements of some of
these objects have been made, and an evident correlation between
linear polarization and stellar brightness was discovered: namely,
the polarized light contribution increased when the star
fades\cite{ber92,grin88,gr94,gri94,grin94}.

Some of the peculiarities observed in the continuum of HAEBE stars
were interpreted by\cite{gri88}. In accordance with the specified
model a star dips into the optically thin disk-like gaseous
envelope. The opaque dust inhomogeneities move around the central
body along elongated orbits. When the star is eclipsed by a dust
cloud the observed brightness decreases and the polarization
increases. In the deep minima, i.e. when the relative contribution
of the scattered light of the gaseous envelope predominates, the
color indices decrease and the star becomes bluer.

The strong variability both of the profiles and the emission lines
intensities also take place in the spectra of HAEBE stars
\cite{gar77,kol77,str72,zai73}. More recently it was argued
\cite{gri95} that kinematic model of the gas-dust envelope
explains  certain details of the complicated transformations in
emission lines. In general terms the following phenomena can
occur. When the dust inhomogeneities move around the star the
ratio of the emission line flux to the flux of the local continuum
(owing to variable screening of the star and the gas envelope) and
the appearance of the profile must change as well.

It is necessary to remember, however, that the available
observational information about the activity of these objects is
still inadequate. Only a few representatives of the assumed HAEBE
stars have been studied in detail. To rectify the existent deficit
of information, to reveal a really homogeneous group (in the
evolutionary sense) of the objects currently identified as HAEBE
stars, and to determine their actual ages, further investigations
with different methods are required. For these reasons
spectroscopic investigations of RR Tau have been carried out.

\section{Observations}
 The spectral observations of RR Tau were carried out from November 1987
 to December 1994 using the slit spectrograph with a
 three-cascades image-tube \cite{den74} attached to the 70-cm
 reflector (Almaty). The spectra were obtained in two spectral
 regions (3800-5100\r{A} and 6200-7600\r{A}) with the dispersion
 20 - 70 \r{A}mm$^{-1}$ and spectral resolution 1.7 - 5.8 \r{A}
 (depending on the dispersion). The spectra were registered on A600 and Kodak OaG films.
 The obtained spectrograms were measured with an automatic Microdensitometer
 and the data were then analyzed using software developed by E.K.Denissyuk.

\section{The results}
The dates of observations, dispersion of spectrograms and values
of the equivalent widths W of the H$_{\alpha}$, H$_{\beta}$\rm\,
and H$_{\delta}$,\rm\,lines are presented in Table 1. The errors
(they were calculated when possible)  characterize deviations from
the average value of W($\lambda$). The error of a single
determination of equivalent widths is about 15-20\%. The behavior
of the H$_{\alpha}$, H$_{\beta}$\rm\, and H$_{\delta}$,\rm\, in
the spectrum of RR Tau is shown in Figures 1-3. We present only
those H$_{\alpha}$ profiles which were obtained with high
dispersion ( about 1.5-2.0\r{A}), apart from two spectra (panels
1,2 of Figure 1). These two spectrograms were obtained with a
spectral resolution about 3\r{A} and they are shown here because
of an interesting detail on the red side of the profile. All
profiles on the Figures 1-3 were corrected for the effects of
limited instrumental contour \cite{bra55}. The intensity axes are
scaled so that the continuum is about unit. The dates of
observations and the exposure time are shown on each panel.

\subsection*{The behavior of H$_{\alpha}$ line}
 It is well known
from high resolution spectroscopic investigations of RR Tau that
the H$\alpha$\rm\, emission line exhibits a double-peaked profile.
The central absorption component reaches the local continuum
level. A variation of the equivalent width W(H$\alpha$)\rm\, by a
factor of approximately 2 has been reported \cite{fin84,gar77}.

Our observations showed variations of W(H$_{\alpha}$)\rm\,up to a
factor of 4-5(see Table 1). The changes in the H$_{\alpha}$\rm\,
profile are clearly seen in Figure 1. As a role a double-peaked
emission profile was observed, where the intensity of blue
component was less than or equal to that of red component. Our
results coincide with those of \cite{zai73}. The case when V/R
ratio is greater than unity \cite{fin84,gar77}  was not seen
during our observations. We can also add that the central
absorption component of the H$_{\alpha}$\rm\, profile reaches the
local continuum level when the spectral resolution of the
spectrograms is adequate.

Red-shifted absorption details were exhibited in the
H$_{\alpha}$\rm\, profile on December 17, 1987. In spite of the
rather low spectral resolution (about 3\r{A}) this peculiarity is
clearly seen on both of the spectrograms that were obtained that
night (Figure 1). Recall that a red-shifted absorption component
of H$_{\alpha}$\rm\, was also observed in the spectrum of the
HAEBE star UX Ori \cite{grin94,kol77,zai73}. Kolotilov
\cite{kol77} remarked that the form of the registered absorption
detail coincided with that of the absorption line of the spectrum
of a standard star, so it might be supposed that the red-shifted
absorption, observed in the UX Ori spectrum, was merely the result
of superposition of H$_{\alpha}$\rm\, emission and stellar line.
On the other hand a shaper and more symmetrical red-shifted
absorption was also registered at H$_{\alpha}$\rm\, in the UX Ori
spectrum. This peculiarity (an inverse P Cygni-like profile )
occurs when the red emission peak disappears \cite{grin94}. It is
interesting that the signature of the inverse P Cygni profile has
been seen in the H$_{\alpha}$\rm\, emission line in the spectrum
of the colder M0V star IP Tau \cite{har87}. This star is the most
likely representative of the T Tau class of objects \cite{her88}.
Consequently, the presence of the inverse P Cygni-like profile at
H$_{\alpha}$\rm\, is a common enough phenomenon for certain young
stars.

\subsection*{Variability of H$_{\beta}$ line}
 Strom et al. \cite{str72}
reported that the H$\beta$\rm\, line in RR Tau spectrum exhibited
weak emission. We registered perceptible activity in the
H$\beta$\rm\, spectral line (Figure 2). At first we see a single
emission peak which overlaps the stellar absorption line. Then
during some nights the only photospheric line was observed. An
emission appears anew on October 30, 1994, and one could see the
remarkable transformation at H$\beta$\rm\, for some days: namely,
the emission peak shifted to short wavelengths on November 7, 1994
and the red absorption detail strengthened. An analogous
phenomenon has been discovered by \cite{kol77} in the spectrum of
UX Ori. A weak H$\beta$\rm\, emission was discerned on some
spectrograms of that HAEBE star, but the emission distorted only
the blue wing of the stellar line, the red wing remained without
any change. It needs to be emphasized that the blue-shifted
H$\beta$\rm\, emission in the spectrum of RR Tau on November 7,
1994 was stronger (Figure 2).\\

\subsection*{The H$_{\delta}$\rm\, line }
 Strom et al. \cite{str72} reported that
shell-like cores have been registered at the higher members of the
Balmer series in RR Tau spectrum. Our observations show rather
symmetrical absorption at H$_{\delta}$. Any emission does not
leave a trace (Figure 3)

\section{Conclusions}
The RR Tau spectroscopic observations allow us to follow up the
behavior the H$_{\alpha}$\rm\, and H$_{\beta}$\rm\, lines in
detail. The ratio of the intensity of the blue component to that
of the red one in the double-peaked H$_{\alpha}$\rm\, profile
varies from 0.6 up to 1.0. The value of W(H$_{\alpha}$) varies
through a rather wide range. An inverse P Cygni-like profile
occurred in the H$_{\alpha}$\rm\, during one night.

We observed wide stellar absorption in the H$_{\beta}$\rm\, line.
A peculiar emission sometimes appeared and overlapped the
absorption line, but to trace the behavior of both the
H$_{\alpha}$\rm\, and H$_{\beta}$\rm\, profiles in more details,
further simultaneous observations of them are needed.

We have compared our results for RR Tau with the available
observational information for UX Ori. UX Ori was investigated in
detail with spectroscopic technique, and one can see that the
profiles of H$_{\alpha}$\rm\, and H$_{\beta}$\rm\, lines in the
spectra of RR Tau and UX Ori change as a whole in a similar
manner. RR Tau and UX Ori are classified as HAEBE-type objects
with large amplitude ( a few magnitudes) of light variations. The
behavior of both the brightness and linear polarization are very
similar for these stars. Hence one can suppose that RR Tau and UX
Ori are really similar in nature.\\

\subsection*{Acknowledgements}
We thank E.K.Denissyuk for allowing us to use his original
software for development of our results and V.N.Gaisina for
measurements of the spectrograms with the Automatic
microdensitometer.

\begin{table}

\begin{center}
Table 1. Some characteristics of spectrograms and equivalent
widths of lines
\end{center}
\begin{center}
\begin{tabular}{rccrccrcc}
\hline
244... & D(\r{A}\rm\,mm$^{-1}$)&W($\lambda$)&244... & D(\r{A}\rm\,mm$^{-1}$)&W($\lambda$)&244... & D(\r{A}\rm\,mm$^{-1}$)&W($\lambda$)\\
 \hline
 \bf{H$\alpha$}&& & 8924,4& 39&40$\pm$4&8278.2&62&-11\\
7126.4& 49 & 64$\pm$6    & .4    & 39&             &.2    &62        &         \\
    .4& 49 &             & .4    & 39&             &8920.4&61        &-8$\pm$2 \\
7147.2& 54 & 31          & 8985.2& 40& 63.3$\pm$3  &    .4&61        &         \\
7238.2& 38 & 73          & 8985.2& 40&             &9556.3&36        &  17     \\
7444.5&42  &73$\pm$10    &9250.4 & 32&  52         &9662.4&38        &   4     \\
    .5& 42 &             &9311.3 & 22&  54         &9664.4&38        &   3     \\
7445.3& 16 &105$\pm$1    &9357.2 & 22& 36$\pm$2    &      \bf{H$\delta$}&&   \\
    .4&16  &             &    .2 & 22&             &7447.4&67        & -11    \\
8271.2& 64 & 52$\pm$3    &9359.2 & 22& 34$\pm$2    &7448.4&68        &-12$\pm$1\\
    .2& 64 &             &     .2& 22&             &    .4&68&\\
8566.4&62  & 115         & 9365.2& 22& 63        &      .4& 68&\\
8569.3&62  &96$\pm$11    &9369.2& 22& 89 &7473.3&62&-11$\pm$1\\
    .3&62  &             &9656.3 & 31& 150& .3& 62&\\
    .3&24  &             &9662.4 &32& 81& .3&62&\\
8573.4& 25 &125& 9664.3&32&86&.3&62&\\
8635.3&62&26$\pm$6&  \bf{H$\beta$}&&&.4&62&\\
    .3&62&   &7447.4&67&-5$\pm$0.4&.4&62&\\
    .3&62&  &.4& 67& & 8278.2& 62&-16$\pm$1\\
    .3&62&  &7448.4&68&-6$\pm$1&.2&62&\\
8918.3&64&22$\pm$4&.4&68&&8920.4&61&-13$\pm$1\\
    .4&25&  &7473.4&62&-4&.4&61&\\
 \hline\\

\end{tabular}
\end{center}
\end{table}
\end{document}